\documentclass[aps,pre,showpacs,twocolumn,superscriptaddress]{revtex4}
\usepackage{graphicx,color,epsfig}
\begin{document}

\title{Spontaneous ordering against an external field in nonequilibrium systems}
\author{J. C. Gonz\'alez-Avella}
\affiliation{IFISC Instituto de F\'isica Interdisciplinar y
Sistemas Complejos (CSIC-UIB), E-07122 Palma de Mallorca, Spain}
\author{M. G. Cosenza}
\affiliation{Centro de F\'isica Fundamental, Universidad de Los
Andes, M\'erida, M\'erida 5251, Venezuela}
\author{V. M. Egu\'iluz}
\affiliation{IFISC Instituto de F\'isica Interdisciplinar y
Sistemas Complejos (CSIC-UIB), E-07122 Palma de Mallorca, Spain}
\author{M. San Miguel}
\affiliation{IFISC Instituto de F\'isica Interdisciplinar y
Sistemas Complejos (CSIC-UIB), E-07122 Palma de Mallorca, Spain}
\date{\today}

\begin{abstract}
We study the collective behavior of nonequilibrium systems subject to an external field with a
dynamics characterized by the existence of non-interacting states. Aiming at exploring the generality of the results, we consider
two types of models according to the nature of their state variables: (i)
a vector model, where interactions are proportional to the overlap between the states, and
(ii) a scalar model, where interaction depends on the distance between states. In both cases the system displays three phases:
two ordered phases, one parallel to the
field, and another orthogonal to the field; and a disordered phase.
The phase space is numerically characterized for each model in a fully connected network.
By placing the particles on a small-world network, we show that, while a regular lattice favors
the alignment with the field, the presence of long-range interactions promotes the formation of the ordered
phase orthogonal to the field.
\end{abstract}


\maketitle

A rather general question considered in the
framework of statistical physics of interacting particles (particles,
spins, agents) is the competition between local
particle-particle interactions (collective self-organization) and
particle interaction with a global externally applied field or
with a global mean field \cite{Lima98,JC2}. Common wisdom answer
to this question is that a strong external field dominates over
local particle-particle interactions and orders the system by
aligning particles with the broken symmetry imposed by the field.
However, this is essentially an equilibrium concept which is not
generally valid for generic non-potential interactions.

In the context of studies of collective phenomena in general systems of
interacting particles, including biological and social systems,
new forms of particle-particle and particle-field interactions
are being considered. There are forms of interaction for which it
turns out that a sufficiently intense external field induces
disorder in the system \cite{JC2,Mazzoni,JC1},  in contrast with the
behavior in, for example, Ising-type systems. Other
intriguing dynamical phenomenon is the collectively
ordering in a state different from the one preferred by the forcing
field. The external field might break symmetry in a given
direction, but the system orders, breaking symmetry in a different
direction. In this paper we examine this situation showing that
these phenomena happen in two recently well studied
non-equilibrium models
\cite{Axelrod97,Marsili,Castellano,Deffuant00,Weisbuch02}. What is
common to these two models is that the particle-particle
interaction rule is such that no interaction exists for some
relative values of the states characterizing the particles that
compose the system. A subsidiary question is the dependence of
this phenomenon on the topology of the network of interactions. We
show that the phenomenon is not found for particles interacting with
its nearest neighbors in a regular lattice, but it occurs in a
globally coupled system: it emerges as long range links in the
network are introduced when going from the regular lattice to a
random network via small world network
\cite{Watts}.

\textit{The vector model}, based in the dynamics of cultural dissemination of Axelrod model, consists of a set of
$N$ particles located at the nodes of an interaction network. The
state of particle $i$ is given by a $F$-component vector $C_i^f$
$(f=1,2,\ldots,F)$ where each component can take any of $q$
different values $C_i^f \in \left\{0,1,\ldots,q-1 \right\}$,
leading to $q^F$ equivalent states \cite{Axelrod97}. The external field, defined as a $F$-component vector $M^f \in \left\{
0,1,\ldots,q-1\right\}$, can interact with any of the particles in
the system.

Starting from a random initial condition, at any given time, a randomly
selected a particle can either interact with the external field or with one of its neighbors.
The dynamics of the system is defined by iterating the following
steps:

\begin{enumerate}
\item Select at random an particle $i$.
\item Select the {\em source of interaction}: with probability
    $B$ the particle $i$ interacts with the field, while with
    probability $(1-B)$ it interacts with one of its nearest
    neighbors $j$.
\item The overlap between the selected particle and the source of
    interaction is the number of shared components between
    their respective vector states, $d =\sum_{f=1}^F
    \delta_{C_i^f,X^f}$, where $X^f = M^f$ if the source of
    interaction is the field, or $X^f = C_j^f$ if $i$ interacts with $j$. If $0<d<F$, with
    probability $d/F$, choose $h$ randomly such that
    $C_i^h\neq X^h$ and set $C_i^h=X^h$.
\end{enumerate}

The strength of the field is represented by the parameter $B \in [0,1]$ that measures the probability for the particle-field
interactions. In the absence of an external field, $B=0$, the system reaches a
stationary configuration in any finite network, where for any pair
of neighbors $i$ and $j$, $d(i,j)=0$ or $d(i,j)=F$. A {\em domain} is a set of connected particles with
the same state. An homogeneous or
ordered phase correspond to $d(i,j)=F$, $\forall i,j$, and
obviously there are $q^F$ equivalent configurations of this state.
An inhomogeneous or disordered phase consist of the coexistence of
several domains. In order to characterize the ordering properties
of this system, we consider as an order parameter the normalized
average size of the largest domain $S$ formed in the system. In finite networks
the dynamics displays a critical point $q_c$ that separates two phases: an
ordered phase ($S \simeq 1$) for $q< q_c$, and a disordered phase
($S \ll 1$) for $q> q_c$ \cite{Marsili,Klemm1,Klemm2,Fede}.

\begin{figure}[t]
\begin{center}
\includegraphics[width=1.\linewidth,angle=0]{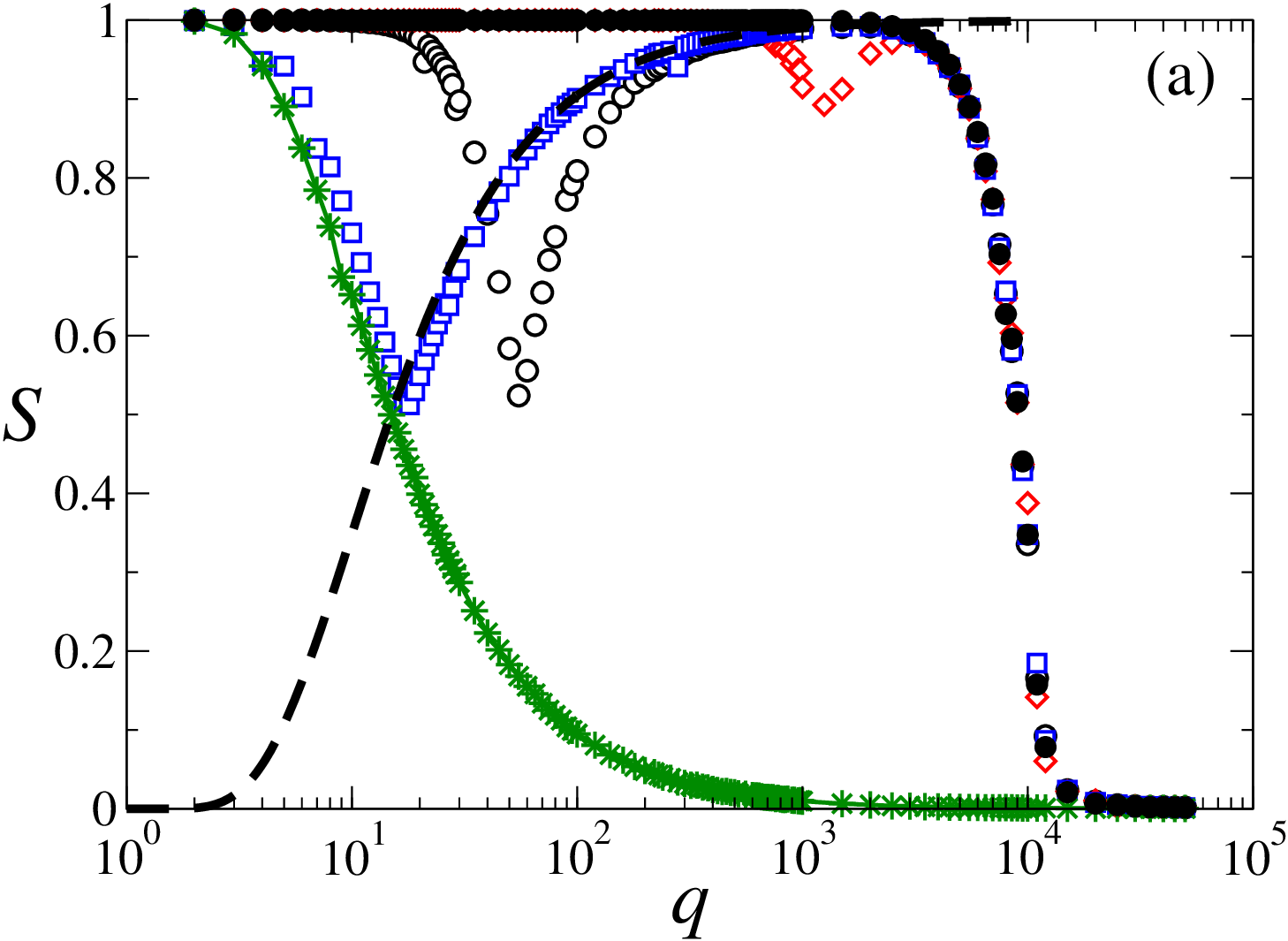}
\end{center}
\vspace{3mm}
\begin{center}
\includegraphics[width=1.\linewidth,angle=0]{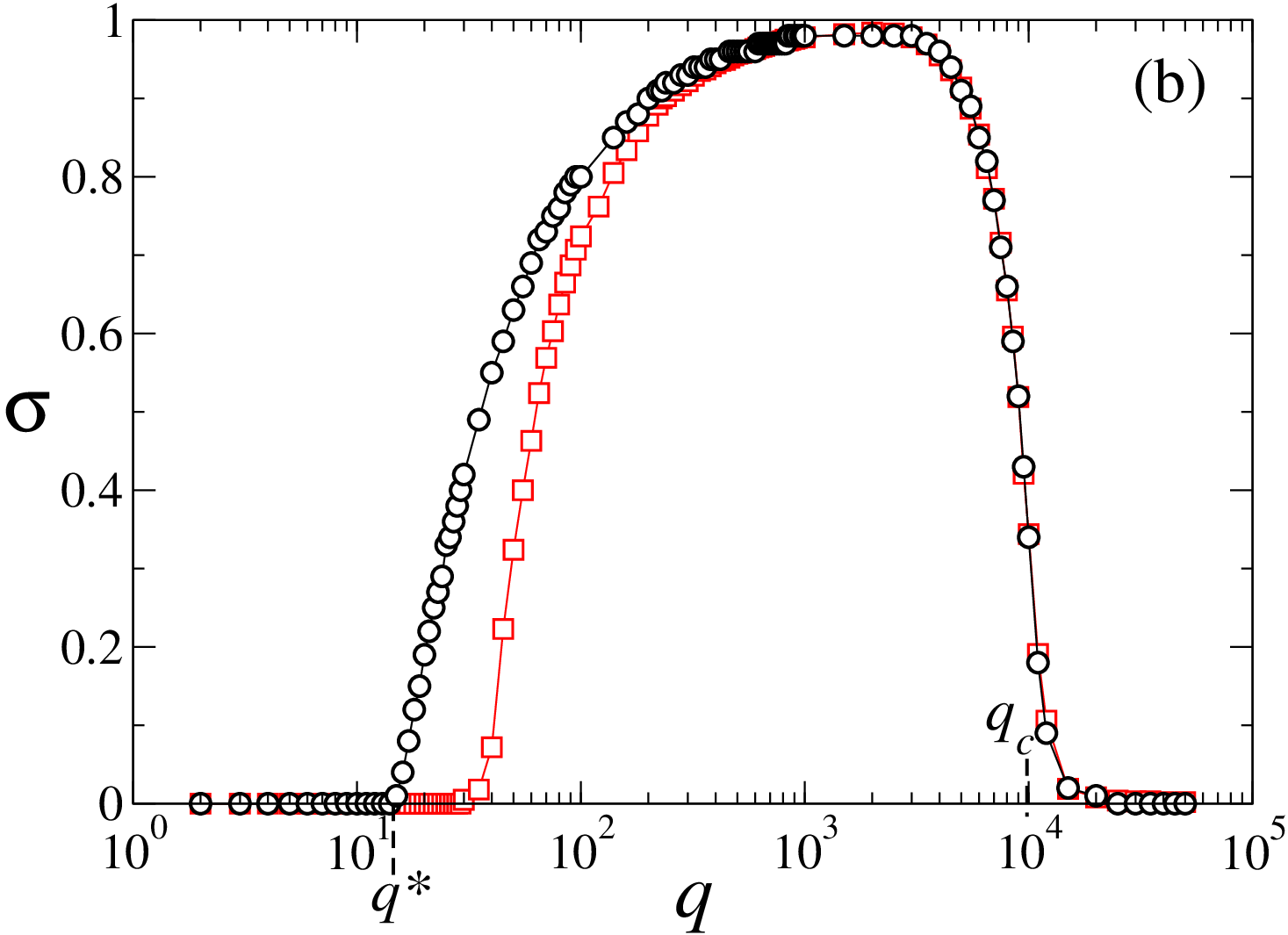}
\end{center}
\caption{ (a) $S$ as a function of $q$ for the Axelrod model on a
fully connected network for $B=0$ (solid circles); $B=0.005$
(diamonds); $B=0.05$ (empty circles); $B=0.5$ (squares); $B=1$
(stars). The continuous line is the analytical curve
$1-(1-1/q)^F$, while the dashed line corresponds to the curve
$(1-1/q)^F$. (b) $\sigma$ versus $q$ for $B=0.8$ (circles) and
$B=0.1$ (squares).
The values of $q_c$ and $q^*$ are indicated for $B=0.8$. Parameter
values are
$N=2500$, $F=10$.}
\end{figure}

First, we analyze the model in a fully connected network.
In the absence of field, i.e. $B=0$, the system spontaneously reaches a
homogeneous state for values $q<q_c \approx 10^4$ (Fig. 1-a ). For $B \to 0$ and  $q<q_c$,
the field $M^f$ is able to impose this homogeneous state to the system,
as in a two-dimensional network \cite{JC1}.
For $B=1$, the particles only interact with
the external field; in this case only those particles that initially
share at least one component of their vector states with the
components of $M^f$ will converge to the field state $M^f$. The fraction
of particles that do not share any component with $M^f$ is given by
$(1-1/q)^F$; thus the fraction of those particles that converge
to  $M^f$ is $1-(1-1/q)^F$. Figure~1 a shows
both the numerically calculated values of $S$ as well as the analytical curve of $S_M$ versus $q$, for
fixed $B=1$. Both quantities agree very well, indicating that the largest domain in the system
possesses a vector state equal to that of the external field when $B=1$.

For intermediate values of
$B$, the spontaneous order emerging in the system for parameter
values $q<q_c$ due to the particle-particle interactions competes with
the order being imposed by the field.
This competition is manifested
in the behavior of the order parameter $S$ which
displays a sharp local minimum at a value $q^*(B)<q_c$ that
depends on $B$, while the value of
$q_c$ is found to be independent of the intensity $B$, as shown in Fig.~1 a.
To understand the nature of this minimum, we plot
in Fig.~1 b the quantity $\sigma=S-S_M$, as a function of $q$, where $S_M$ is the normalized average size of the largest
domain displaying the state of the field $M^f$. For $q <
q^*(B)$ the largest domain corresponds to the state of the
external field, $S = S_M$, and thus $\sigma=0$. For $q>q^*(B)$, the
largest domain no longer corresponds to the state of the external
field $M^f$ but to other state non-interacting with the external
field, i.e., $S>S_M$, and $\sigma > 0$. The value of $q^*(B)$ can
be estimated for the limiting case $B \to 1$, for which
$S_M \approx 1-(1-1/q)^F$ and the largest domain different from the field is $S \approx 1-S_M$.
Therefore the condition $S=S_M$  yields
$q^*(B \to 1)=\left[ 1-(1/2)^{1/F}\right]^{-1}$. For $F=10$ it gives $q^*(B \to 1)=15$
in good agreement with the numerical results.
The order parameter $\sigma$ reaches a
maximum at some value of $q$ between $q^*$ and $q_c$ above which
order decreases in the system and both $S \rightarrow 0$, $S_M
\rightarrow 0$. As a consequence, $\sigma$ starts to decrease.

\begin{figure}[t]
\includegraphics[width=1\linewidth,angle=0]{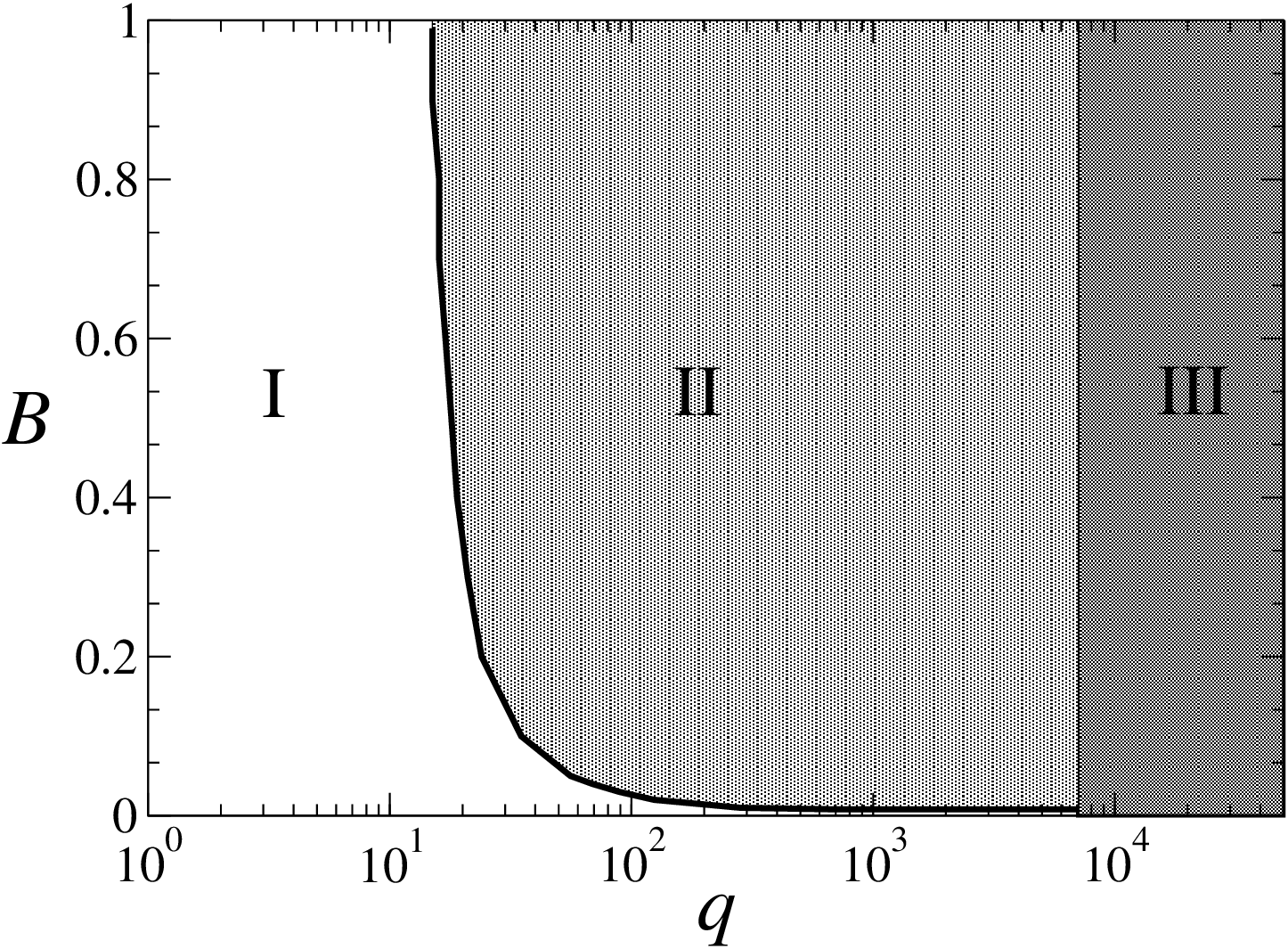}
\caption{Phase space on the plane $(q,B)$ for the vector model
on a fully connected network subject to an external field, with fixed
$F=10$.
Regions where the phases I, II, and III occur are indicated.}
\end{figure}

The collective behavior
of the vector model on a fully connected network subject to an
external field can be characterized by three phases on the space of parameters $(q,B)$, as shown in Fig.~2: (I) an
ordered phase induced by the field for $q < q^*$, for
which $\sigma=0$ and $S=S_M\neq 0$; (II) an ordered
phase in a state different from that of the field for $q* <q< q_c$, for which $\sigma$ increases and $S>S_M$;
and (III) a disordered phase for $q >q_c$, for which $\sigma$
decreases and  $S \rightarrow 0$, $S_M \rightarrow 0$.

For parameter
values $q<q_c$ for which the system orders due to the interactions
among the particles,
a sufficiently weak external field is able to impose its state to the entire system (phase I). However, if the probability of interaction with the field $B$
exceeds a threshold value, the system spontaneously orders in a state different from that of the field (phase II).

\textit{Continuous states} based on bounded
interactions provide other instances of a nonequilibrium systems
where induced and spontaneous order compete in the presence of an
external field. Consider, for example, the bounded confidence
model \cite{Deffuant00}. It consists of a population of $N$ particles
where the state of particle $i$ is given by a real number $C_i \in
[0,1]$. We introduce an external field $M \in [0,1]$ that can
interact with any of the particles in the system. The strength of the field
is again described by a parameter $B \in [0,1]$ that measures the probability
for the particle-field interactions, as in the vector model.

\begin{figure}[t]
\begin{center}
 \includegraphics[width=1\linewidth,angle=0]{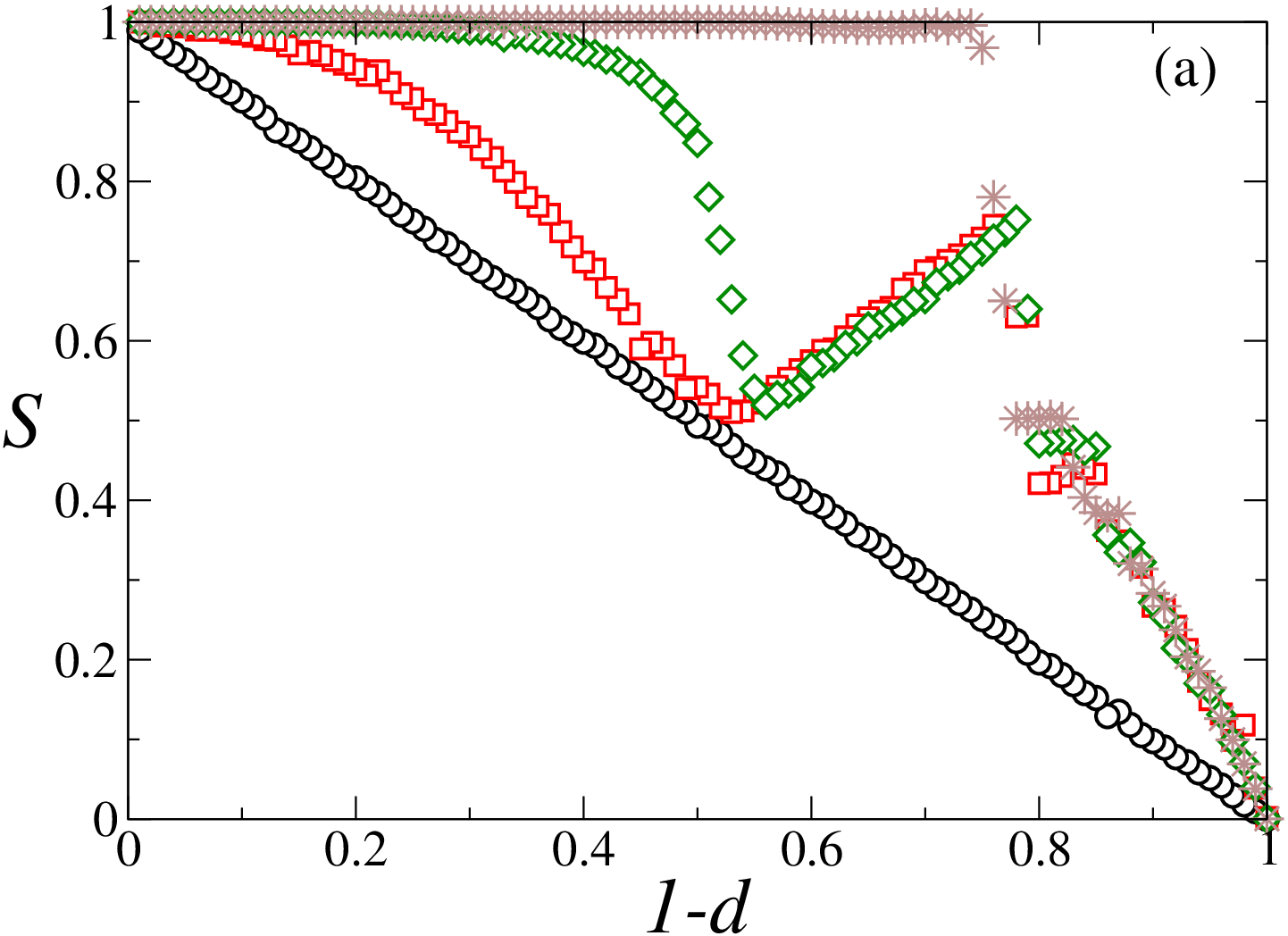}
\end{center}
\vspace{5mm}
\begin{center}
\includegraphics[width=1\linewidth, angle=0]{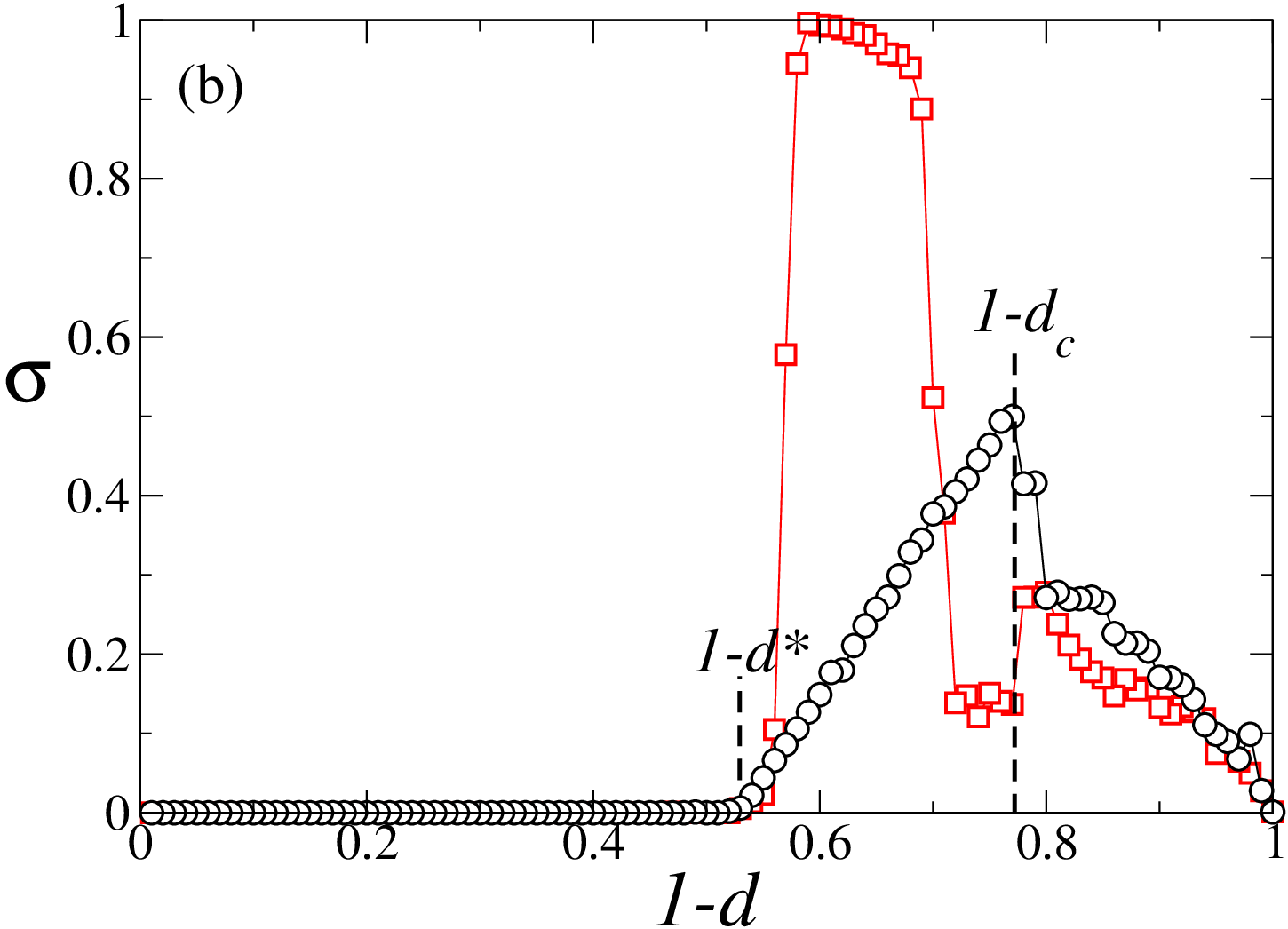}
\end{center}
\caption{ (a) $S$ versus $1-d$ for the continuous model for $B=0$ (stars);
$B=0.5$ (diamonds); $B=0.8$ (squares); $B=1$ (circles). (b)
$\sigma$ vs. $d$ for $B=0.8$ (circles) and $B=0.1$ (squares). The values
of $1-d_c$ and $1-d^*$
are indicated for $B=0.8$. Size of the system is $N=2500$.}
\end{figure}

We start from a uniform, random initial distribution of the states
of the particles.  At each time step, an particle $i$ is randomly chosen;
\begin{enumerate}
\item with probability $B$, particle $i$ interacts with the field
    $M$: if $|C_i-M|<d$, then
    \begin{equation}
     C_i^{t+1} =  \frac{1}{2} (M+C^{t}_i)~;
    \end{equation}
\item otherwise, a nearest neighbor $j$ is selected at random: if
    $|C_i-C_j| < d$ then:
    \begin{equation}
    C^{t+1}_i =  C^{t+1}_j = \frac{1}{2} (C^{t}_j+C^{t}_i)~.
    \end{equation}
\end{enumerate}
The parameter $d$ defines a threshold distance for interaction and the remainder we
fix $M=1$.

\begin{figure}[h]
\includegraphics[width=1\linewidth,angle=0]{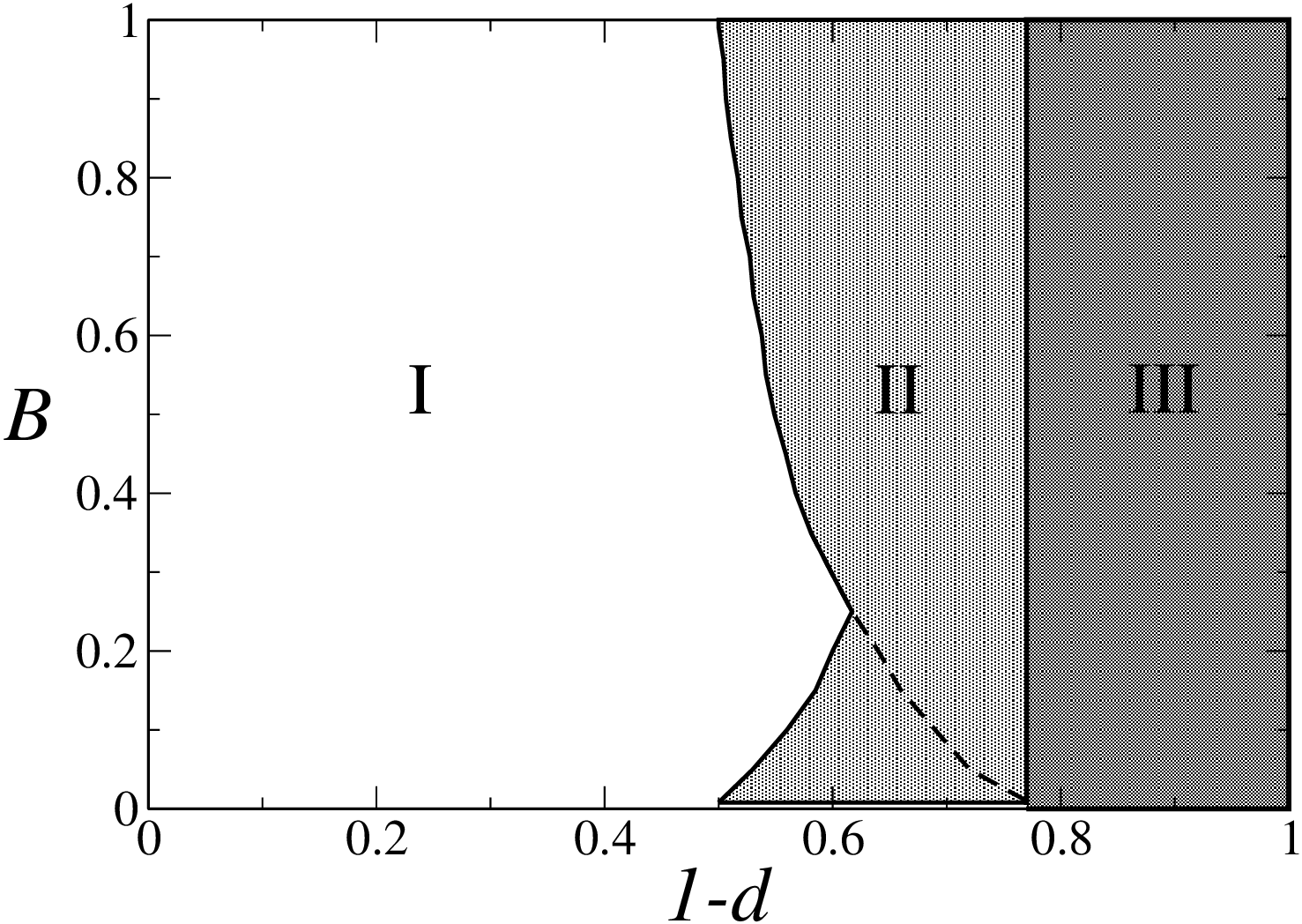}
\caption{Phase space on the plane $(1-d,B)$ for the scalar model
on a fully connected network subject to an external field. Regions
where the phases I, II, and III occur are indicated. The dashed line in phase II separates region where the maximum of $\sigma \rightarrow 1$ (below this line) from the region where $\sigma \leq 0.5$ (above this line).
}
\end{figure}

We calculate the normalized
average size of the largest domain $S$ in the system as a function of $1-d$,
for different values of $B$, as shown in Fig.~3(a). For $B=0$, the system spontaneously
reaches a homogeneous state $C_i=0.5$, $\forall i$, characterized
by $S=1$, for values $1-d<1-d_c \approx 0.77$, with $d_c \approx 0.23$ \cite{Deffuant00};
while for $1-d>1-d_c$ several domains are formed yielding $S<1$.

For $B=1$ particles only interact with the field; in this case the
value of $M$ is imposed on the largest domain whose normalized
size increases with the threshold, i.e. $S=d$.  For intermediate values of
$B$, the spontaneous order emerging in the system for values of $1-d<1-d_c$
due to the interactions between the particles competes with the order being induced by the field.
The quantity $S$ exhibits a sharp local minimum at a value $1-d=1-d^*<1-d_c$, as shown in
Fig.~3(a). In Fig.~3(b) we plot the order parameter $\sigma=S-S_M$ as a function
of $1-d$, for different values of $B$. For $1-d <1-d^*$ the largest domain reaches a state equal to $M$, that
is $S=S_M$, and thus $\sigma=0$. At $1-d=1-d^*$, the state of the
field no longer corresponds to the largest domain, i.e., $S>S_M$,
and $\sigma$ starts to increase as $1-d$ increases.
For a small value of $B$, the quantity $\sigma$ reaches a maximum close to one, indicating that
the spontaneously formed largest domain almost occupies the entire system, i.e., the field is too weak to compete with the attracting homogeneous state $C_i=0.5$, $\forall i$. However, when $B$ is increased, the maximum of
$\sigma$ is about $0.5$, i.e., the attraction of the field $M=1$ increases and the size of the domain with a state equal to $M$ is not negligible in relation to the size of the largest domain. In contrast, in the vector model the maximum $\sigma \rightarrow 1$ in the region $q* <q< q_c$, independently of the value of $B$.

The value of
$d^*$ in the scalar model depends on $B$ and it can be estimated for $B \rightarrow
1$. In this case, $S_M \approx d$ and $S \approx 1-d$; thus the
condition $S=S_M$  yields $d^* \approx 0.5$ when $B \rightarrow
1$. The quantity $\sigma$ reaches a maximum at the value $1-d \approx 1-d_c$,
above which disorder increases in the system, and both $S$ and
$S_M$ decrease. As a consequence, $\sigma$ decreases for $1-d>1-d_c$.

As in the vector model, the collective behavior exhibited by the
scalar model on a fully connected network subject to an external
field can be characterized by three phases: (I) an ordered
phase parallel to the field for $1-d < 1-d^*$, for which
$\sigma=0$ and $S=S_M\neq 0$; (II) a ordered phase for
$1-q^* < 1-d < 1-d_c$, for which $\sigma$ increases and $S>S_M$; and
(III) a disordered phase for $1-d > 1-d_c$, for which $\sigma$
decreases and both $S$ and $S_M$ decrease. Figure~4 shows the
phase diagram on the plane $(1-d,B)$ for the scalar model subject to
an external field. The continuous curve
separating phases I and II gives the dependence $d^*(B)$.

\textit{Short range interactions}\textit{}. To analyze the role of the
connectivity on the emergence of an ordered phase orthogonal to the external field,
we consider a small-world network \cite{Watts}, where the rewiring probability can
be varied in order to introduce long-range interactions between the particles.
We start from a two-dimensional lattice
sites with nearest-neighbor interactions. Each connection is
rewired at random with probability $p$.
The value $p=0$ corresponds to a regular network, while
$p=1$ corresponds to a random network with $\langle k \rangle =
4$.

Figure~5 shows the order parameter $S$ as a function of $q$ in the
vector model defined on this network for different values of the
rewiring probability $p$ and for a fixed value of the intensity of
the field $B$. The critical value $q_c$ where the order-disorder transition takes place
increases with $p$, which
is compatible with the large value of $q_c$ observed in a fully
connected network. When the long-range interactions between particles
are not present, i.e. $p=0$, the external field is able to impose its
state to the entire system for $q<q_c$. Spontaneous ordering
different from the state of the external field appears as the
probability of having long-range interactions increases. The size of this alternative largest domain
increases with $p$, but it does not grow enough to cover the
entire system (see inset in Fig.~5). Increasing the rewiring
probability in the scalar model also produces a behavior similar
to the vector model.
Thus, in systems whose dynamics is based on a
bound for interaction, the presence of long-range connections
facilitates the emergence of spontaneous ordering not associated
to the state of an applied external field.

\hspace{1cm}
\begin{figure}[t]
\includegraphics[width=1.0\linewidth,angle=0]{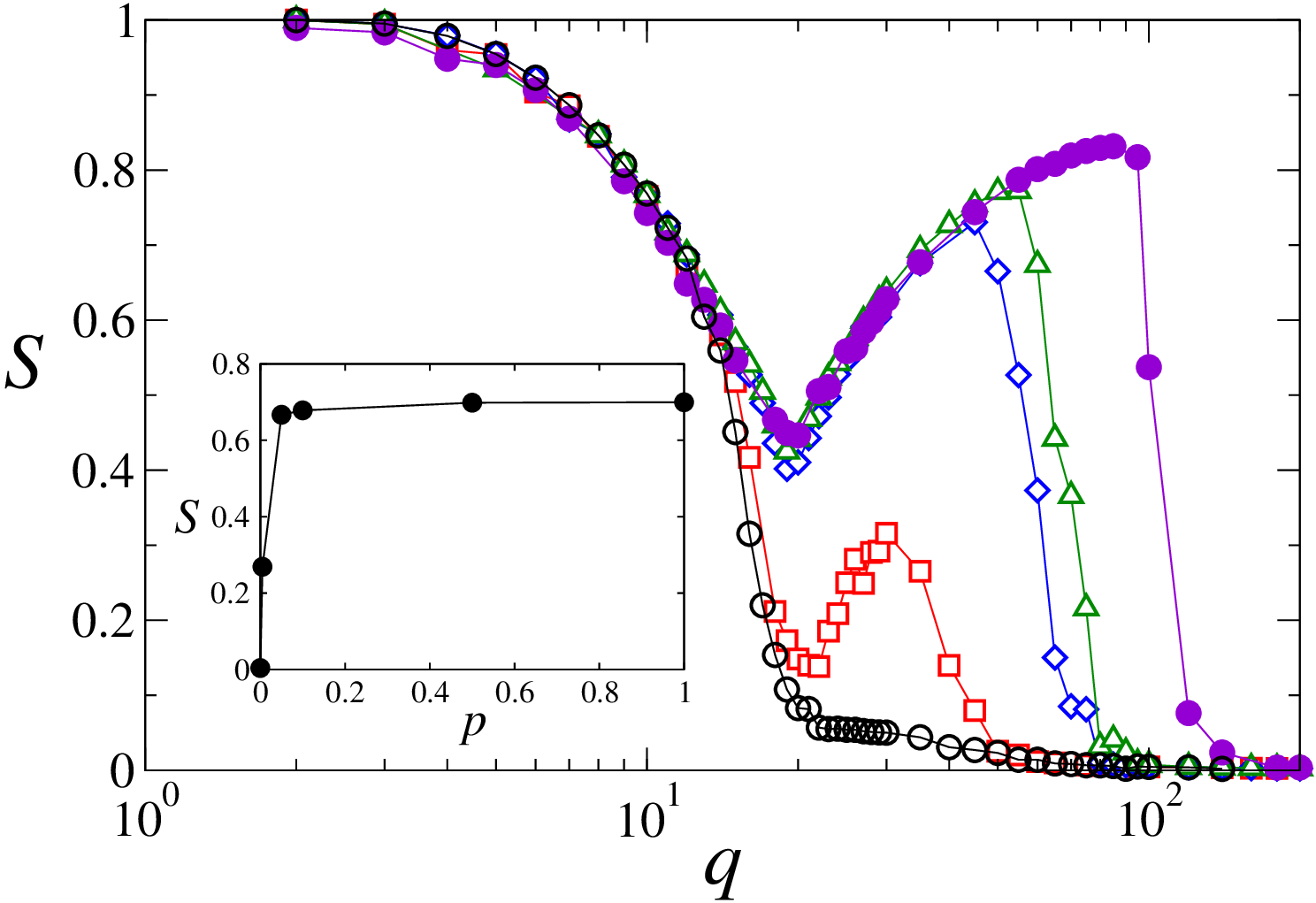}
\caption{$S$ versus $q$ in the vector model on a small world
network with $\langle k \rangle =4$, $B=0.5$, $F=3$, for different
values of the probability $p$: $p=0$ (empty circles), $p=0.005$
(squares), $p=0.05$ (diamonds), $p=0.1$ (triangles), $p=1$ (solid
circles). Inset: $S$ vs. $p$ for fixed values $q = 40 > q^*$ and
$B=0.5$.}
\end{figure}

In summary, we have studied the collective behavior of nonequilibrium
systems with non-interacting states and subject to an external field.
We have considered two models on a fully connected network that share a common feature: the existence of non-interacting states.
In both cases we have found three phases  depending on parameter values:
two ordered phases, one having  a
state equal to the external field, an another ordered phase, consisting of a large domain with a
state orthogonal to the field; and a disordered phase.
The occurrence of an ordered phase with a state orthogonal to the field is
enhanced by the presence of long range
connections in the underlying network. We have verified that this alternative ordered phase also appears when the models considered here are defined on a scale-free network.

The emergence of an ordered phase with a state
different from that of an external field
may be relevant in social systems as well as in many biological systems having motile elements,
such as swarms, fish schools, and bird flocks \cite{Mikhailov},
whose dynamics usually possess a bound condition for interaction. Thus one may expect that this phenomenon
should arise in large class of nonequilibrium systems in the presence of an external source for interaction.

We thank F. Vazquez and K. Klemm for useful discussions. J.C.G-A.,
V.M.E., and M.S.M. acknowledge support from MEC (Spain) through
projects FISICOS FIS2007-60327 (Spain). M.G.C. acknowledges
support from CDCHT, Universidad de Los Andes (Venezuela) under
grant No. C-1579-08-05B.

\end{document}